\documentclass[twocolumn]{aastex701}
\usepackage{xcolor}

\begin{document}

\title{Evolution of Massive Main-Sequence Stars in Binary Population Synthesis. II. Consequences for Black Hole High-Mass X-ray Binaries and Binary Black Holes}

\author[0009-0002-2040-0637]{Adam Br\v{c}ek}
\affiliation{School of Physics and Astronomy, Monash University, Clayton, VIC 3800, Australia}
\affiliation{OzGrav: The ARC Centre of Excellence for Gravitational Wave Discovery, Clayton, VIC 3800, Australia}
\email{adam.brcek@monash.edu}

\author[0000-0002-8032-8174]{Ryosuke Hirai}
\affiliation{Astrophysical Big Bang Laboratory (ABBL), Pioneering Research Institute (PRI), RIKEN, Wako, Saitama 351-0198, Japan}
\affiliation{School of Physics and Astronomy, Monash University, Clayton, VIC 3800, Australia}
\affiliation{OzGrav: The ARC Centre of Excellence for Gravitational Wave Discovery, Clayton, VIC 3800, Australia}
\email{ryosuke.hirai@monash.edu}

\author[0000-0002-6134-8946]{Ilya Mandel}
\affiliation{School of Physics and Astronomy, Monash University, Clayton, VIC 3800, Australia}
\affiliation{OzGrav: The ARC Centre of Excellence for Gravitational Wave Discovery, Clayton, VIC 3800, Australia}
\email{ilya.mandel@monash.edu}

\author[0009-0000-2208-5429]{Jack Yutong Chen}
\affiliation{School of Physics and Astronomy, Monash University, Clayton, VIC 3800, Australia}
\affiliation{OzGrav: The ARC Centre of Excellence for Gravitational Wave Discovery, Clayton, VIC 3800, Australia}
\affiliation{Department of Physics, University of Auckland, Private Bag 92019, Auckland, New Zealand}
\email{ych993@aucklanduni.ac.nz}

\correspondingauthor{Adam Br\v{c}ek}
\email{adam.brcek@monash.edu}

\begin{abstract}

Interactions and mass loss during the main-sequence (MS) phase of massive binary stars can significantly alter their subsequent evolution and the properties of the black holes (BHs) they ultimately produce. We use the rapid binary population synthesis code \textsc{COMPAS}, incorporating a new treatment of MS evolution that self-consistently tracks the convective core under arbitrary mass-loss or mass-gain histories. We explore the properties of black hole high-mass X-ray binaries (BH-HMXBs) and binary black holes (BBHs) formed under this new framework. We find that a more detailed treatment of MS evolution is necessary to reproduce the high BH masses observed in BH-HMXBs, which are otherwise absent under prescriptions that do not take the internal structure of MS stars into account. Matching the observed rates and orbital periods favors case A mass transfer that is more non-conservative than commonly assumed. We further find that mass transfer of helium-enriched material during case A mass transfer is common in massive binaries, with potential consequences for the structure of the accreting star. For the BBH population, the new MS framework alters the chirp-mass and mass-ratio distributions of detectable mergers, and case A mass transfer can make a substantial contribution to the BBHs formed through the stable mass transfer channel. These systems show a strong preference for near-equal mass ratios, broadly consistent with recent constraints from gravitational-wave observations.

\end{abstract}

\keywords{\uat{Binary stars}{154} --- \uat{Main sequence stars}{1000} --- \uat{High mass X-ray binary stars}{733} --- \uat{Black holes}{162}}


\section{Introduction} \label{sec:intro}
In a massive binary system, interactions during the main-sequence (MS) phase, while the stars are still core-hydrogen burning, strongly influence the binary's subsequent evolution and the properties of the black holes (BHs) that may form. These early interactions can alter stellar masses, spins, internal structure, and orbital parameters, leaving imprints that can be observed during later evolutionary stages. The properties of binaries containing BHs observed today therefore provide important insight into their formation pathways. In particular, black hole high-mass X-ray binaries (BH-HMXBs), systems in which a BH accretes material from a massive stellar companion, offer a unique window into BHs and the binary evolutionary processes that lead to their formation.

The formation and properties of BH-HMXBs are known to be sensitive to both metallicity and the treatment of mass transfer and angular momentum loss \citep{Romero-Shaw2023MNRAS.524..245R, Willcox2023ApJ...958..138W, Xing2025A&A...693A..27X}. Metallicity regulates stellar wind mass loss, affecting the final mass of the BH, while the conservativeness of mass transfer and the angular momentum lost from the binary during non-conservative mass transfer govern the orbital separation and mass-ratio evolution. Because BH-HMXBs originate from massive stars, an accurate treatment of MS evolution is particularly important when modeling these systems. Observations indicate that the majority of massive stars are in binaries or higher-order multiples and are likely to interact while still on the MS, significantly impacting their subsequent evolution \citep{Sana2012Sci...337..444S, Moe2017ApJS..230...15M}. At higher metallicities, strong line-driven winds can remove a significant fraction of the initial stellar mass already during core-hydrogen burning, further modifying the stellar structure and influencing post-MS evolution. 

To date, there are three confirmed BH-HMXBs with dynamically measured BH masses: Cygnus X-1 \citep{Miller-Jones2021Sci...371.1046M, Ramachandran2025A&A...698A..37R}, LMC X-1 \citep{Orosz2009ApJ...697..573O}, and M33 X-7 \citep{Valsecchi:2010,Ramachandran2022A&A...667A..77R}. In all three systems, the donors are massive MS stars, near Roche-lobe filling, and the BHs are inferred to be rapidly spinning (though spin measurements may suffer from significant systematics, see, e.g., 
\citealt{Reynolds:2020} for a review). The high spins of these BHs are thought to be inherited from their progenitor stars rather than produced by accretion onto the BH itself, as achieving such spins through Eddington-limited accretion would require unrealistically large accreted masses given the short lifetimes of massive stellar companions \citep{King1999MNRAS.305..654K}. Instead, early binary interactions, particularly case A mass transfer (when the donor is core hydrogen burning), are expected to play a key role \citep{Qin2019ApJ...870L..18Q, Neijssel2021ApJ...908..118N}. During this phase, orbital hardening and subsequent tidal locking can significantly spin up the BH progenitor while it is still on the MS. Since the stellar core and envelope are not yet decoupled, this allows angular momentum deposited at the surface to be efficiently redistributed throughout the star. This mechanism provides a formation channel for rapidly rotating BHs in tight BH-HMXB systems.

Further insight into the lives of BHs is provided by gravitational-wave observations from the LIGO-Virgo-KAGRA (LVK) detector network, which has now reported 390 compact-object merger candidates with the fifth Gravitational-Wave Transient Catalog (GWTC-5.0, \citealt{LIGO2026arXiv260527225T}). These observations have revealed a diverse population of binary black holes (BBHs), spanning a wide range of masses. The formation pathways of BBHs formed through isolated binary evolution, both for individual systems and at the population level, can be explored using binary population synthesis \citep[][]{Belczynski2008ApJS..174..223B,Breivik2016ApJ...830L..18B,Eldridge2016MNRAS.462.3302E,Stevenson2017NatCo...814906S, Mapelli2019MNRAS.487....2M,Broekgaarden2022MNRAS.516.5737B,Boesky2024ApJ...976...24B,Briel2026arXiv260203629B}. In particular, well-measured parameters such as the chirp mass provide valuable constraints on the underlying stellar and binary physics. A subset of merging BBHs may originate from systems that previously experienced a BH-HMXB phase, linking X-ray binaries and gravitational-wave sources through a common evolutionary history \citep{Gallegos-Garcia2022ApJ...938L..19G, Liotine2023ApJ...946....4L}.

BBHs from isolated binary evolution are generally formed through two main channels: stable mass transfer (SMT) or common-envelope (CE) evolution. In the SMT channel, the mass transfer efficiency and angular momentum loss can significantly impact the outcome of these interactions \citep{Willcox2023ApJ...958..138W,vanSon2022ApJ...940..184V,Picco2024A&A...681A..31P}. Recent studies have also highlighted the importance of case A mass transfer for BBH mergers \citep{Klencki2026A&A...706A.296K,Briel2026arXiv260203629B}, a channel that is often neglected in rapid binary population synthesis models, which oversimplify the evolution of MS stars.

In this work, we investigate the population of BH-HMXBs and BBHs formed using the rapid binary population synthesis code \textsc{COMPAS} with a new MS evolution framework for massive stars. This new approach self-consistently tracks the evolution of the convective core under arbitrary mass-loss and mass-gain histories. The theoretical framework and its implementation in \textsc{COMPAS} are described in detail in \citet{Brcek2026ApJ..1002...78B}. We summarize the population synthesis assumptions, procedure, and post-processing in Section~\ref{sec:method}. We first focus on the population of BH-HMXBs in Section~\ref{sec:results_bh_hmxb}, examining how metallicity and mass transfer assumptions shape their masses and orbital periods, and exploring their alternative formation channels. We then separately focus on the population of BBHs in Section~\ref{sec:results_bbh} and explore how the new MS evolution framework affects both the intrinsic and observed properties of the merging BBH population. We discuss the results and limitations of our models in Section~\ref{sec:discussion}, and summarize the main results in Section~\ref{sec:conclusion}.

\section{Method} \label{sec:method}
We model the evolution of massive binary stars using the rapid binary population synthesis code \textsc{COMPAS}\footnote{Publicly available via \url{https://github.com/TeamCOMPAS/COMPAS}} \citep[Compact Object Mergers: Population Astrophysics and Statistics;][]{Stevenson2017NatCo...814906S, Vigna-Gomez2018MNRAS.481.4009V, Riley2022ApJS..258...34R, Mandel2025ApJS..280...43T}, version 03.29.03. \textsc{COMPAS} follows the evolution of binary stars from the zero-age main sequence (ZAMS) through all subsequent phases of stellar and binary evolution, including mass transfer, common-envelope evolution, supernovae, and compact-object formation. The code's computational efficiency allows the production of large populations of synthetic binaries, making it well suited for exploring uncertainties in binary evolution physics. The single star evolution in \textsc{COMPAS} is based on analytic fits to detailed stellar models, with binary interactions treated using parametrized prescriptions \citep{Pols1998MNRAS.298..525P, Hurley2000MNRAS.315..543H, Hurley2002MNRAS.329..897H}. 

\begin{figure*}[t!]
    \plotone{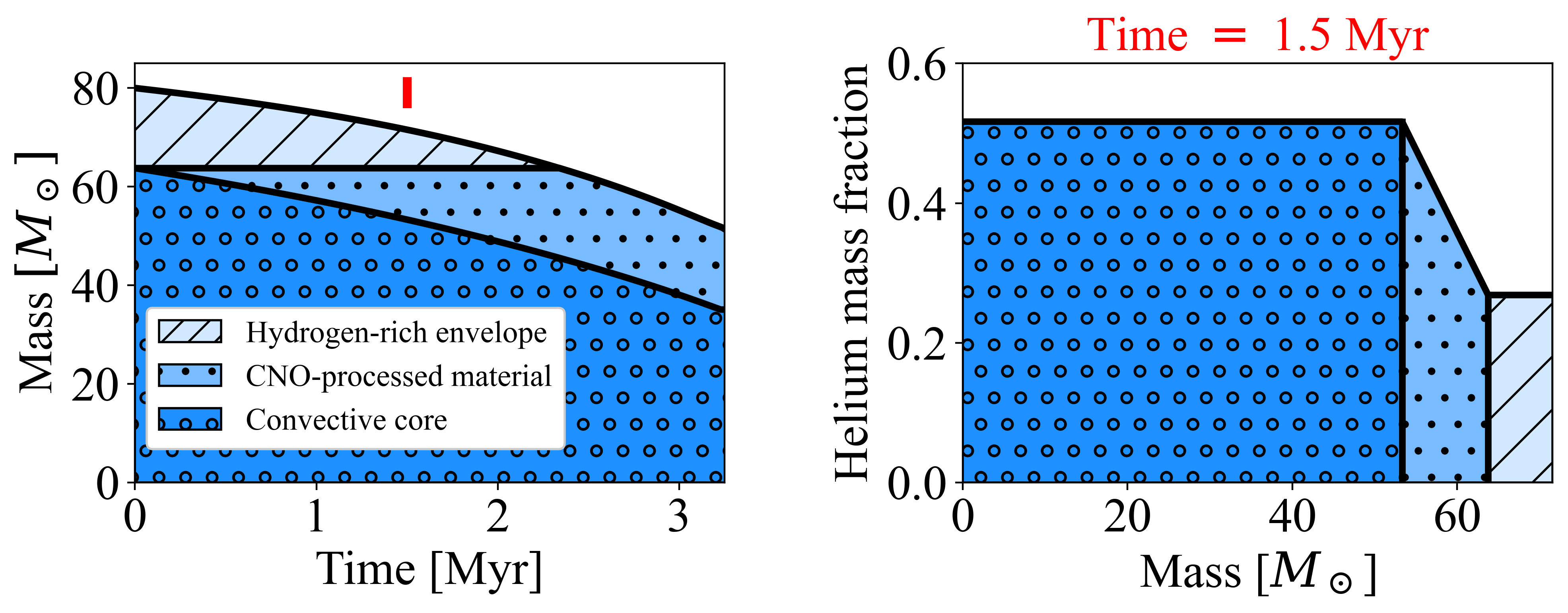}
    \caption{The left panel shows the mass evolution of an initially $80\:M_\odot$ star at solar metallicity over its MS lifetime under the \citet{Brcek2026ApJ..1002...78B} framework, showing the total stellar mass (uppermost boundary) and the extents of the hydrogen-rich envelope (diagonal hatching), CNO-processed material (dots), and convective core (open circles). Once the star is sufficiently stripped by stellar wind mass loss, helium-enriched material from the CNO-processed layers is expected to appear at the surface, enhancing the surface helium abundance (visible here after $\sim2$ Myr). The right panel shows a snapshot of the helium profile at $1.5$ Myr (indicated by the red marker in the left panel), where the shaded areas correspond to the helium mass contained in each layer. The framework assumes a linear transition in helium mass fraction between the convective core and the envelope.}
\label{fig:helium_profile}
\end{figure*}

In this work, we adopt a new treatment of MS evolution introduced by \citet{Brcek2026ApJ..1002...78B}, which is based on detailed stellar models from \citet{Shikauchi2025ApJ...984..149S}. This framework describes the coupled evolution of the convective core mass and central helium abundance, which then affect other stellar properties such as the radius, surface abundances, luminosity, and evolutionary timescales. It also tracks the extent of CNO-processed material within the star, allowing the internal helium and hydrogen profiles to be reconstructed at any point during MS evolution (Figure \ref{fig:helium_profile}). This has been utilized to improve the treatment of surface abundance evolution, rejuvenation of MS accretors, and MS merger products \citep{Brcek2026ApJ..1002...78B}. The new MS evolution framework effectively replaces the evolution tracks of massive stars from \citet{Hurley2000MNRAS.315..543H}, with the exception of the stellar radius, where we adopt the functional forms of the \citet{Hurley2000MNRAS.315..543H} formalism. Throughout this work, we refer to the framework of \citet{Brcek2026ApJ..1002...78B} as the \texttt{BRCEK} MS evolution framework. We refer to the original approach of \citet{Hurley2000MNRAS.315..543H}, in which MS stars are characterized only by their total mass and age, as the \texttt{HURLEY} framework.

To study the impact of the new MS evolution framework on BH populations, we perform a suite of population synthesis simulations with \textsc{COMPAS}. Unless stated otherwise, all other physical assumptions are kept at their default values. These include the preferred prescription for stellar winds from \citet{Merritt2025arXiv250717052M}, the pulsational pair-instability prescription following \citet{Marchant2019ApJ...882...36M}, and the prescription for natal kicks of compact objects following \citet{Mandel2020MNRAS.499.3214M} with the correction from \citet{Disberg2025ApJ...989L...8D}. We adopt the $\alpha$-$\lambda$ formalism for common envelopes \citep{Webbink1984ApJ...277..355W, deKool1990ApJ...358..189D} with $\alpha=1$ and the $\lambda$-prescription from \citet{Xu2010ApJ...716..114X} as revised in \citet{Mandel2025ApJS..280...43T}. The stability of mass transfer is determined by comparing the response of the donor's radius to mass loss against the response of its Roche lobe. For MS and Hertzsprung-gap (HG) donors, we adopt fixed values for the adiabatic stellar radius response following \citet{Vigna-Gomez2018MNRAS.481.4009V}, while for evolved giants we follow \citet{Soberman1997A&A...327..620S}.

In the default \textsc{COMPAS} settings, the accretion rate during Roche-lobe overflow is limited by the thermal (Kelvin-Helmholtz) timescale of the accretor \citep{Hurley2002MNRAS.329..897H}. The accretor can absorb mass at up to 10 times its own thermal rate, so the mass transfer efficiency is determined as
\begin{equation}\label{eq:thermal}
    \beta_\mathrm{thermal} = \min \left( 10\times\frac{M_\mathrm{accretor}}{\dot M_\mathrm{donor}\:\tau_\mathrm{accretor}}, \:1 \right) ,
\end{equation}
where $\dot M$ is the mass loss rate, $M$ is the total stellar mass, and $\tau$ denotes the thermal timescale. The mass transferred in excess of this rate is lost from the system. When mass transfer is non-conservative, we assume that the excess mass is lost isotropically, carrying away the specific angular momentum of the accretor. Accretion onto compact objects is assumed to be Eddington-limited. 

Initial binary parameters at ZAMS are drawn from probability distributions. The mass of the primary (initially more massive) star is sampled from the \citet{Kroupa2001MNRAS.322..231K} initial mass function in the mass range of $5$--$150$ $M_\odot$. The mass ratio is drawn from a uniform distribution \citep{Sana2012Sci...337..444S}, with a minimum secondary mass of 0.1 $M_\odot$. The orbital separation is drawn from a log-uniform distribution in the range of $0.1$--$1000$ au \citep{Opik1924PTarO..25f...1O, Sana2012Sci...337..444S}, and all systems are initialized with zero eccentricity. In our simulations, we vary the MS evolution formalisms (\texttt{BRCEK} vs. \texttt{HURLEY}), the metallicity $Z$, and the mass transfer efficiency $\beta$.

We classify the formation channels according to the evolutionary phase of the primary at the onset of its first mass transfer episode, following the stellar type definitions of \citet{Hurley2000MNRAS.315..543H}. Case A mass transfer refers to systems in which the primary initiates mass transfer while still on the MS; this includes case AB mass transfer, in which the donor evolves off the MS and onto the Hertzsprung gap (HG) over the course of the interaction. Case B mass transfer refers to systems in which the primary initiates mass transfer during the HG, first giant branch, or core-helium-burning phase. Case C mass transfer refers to systems in which the donor initiates mass transfer after the end of core helium burning. 

For the simulations of BH-HMXBs, we evolve binaries at three fixed metallicities: $Z_\odot$, $Z_\odot/2$, and $Z_\odot/10$, where $Z_\odot=0.0142$ \citep{Asplund2009ARA&A..47..481A}. We use the default remnant mass prescription from \citet{Mandel2020MNRAS.499.3214M}. To assess the sensitivity of the population to mass transfer physics, we additionally perform simulations with fixed mass transfer efficiencies of $\beta=0$ (fully non-conservative mass transfer), 0.5, and 1 (fully conservative). This represents the fraction of the mass accreted by the companion and is applied to all mass transfer episodes during the evolution of a binary, except mass transfer onto compact remnants.

For each variation, we evolve $5\times10^5$ binaries. BH-HMXB candidates are identified using the observability criterion for wind-accreting X-ray binaries from \citet{Hirai2021PASA...38...56H}. We select as BH-HMXB candidates the systems containing a BH and a MS companion that is at least 80\% Roche-lobe filling.

To estimate the number of BH-HMXBs observable in the Milky Way at the present day, we follow the approach of \citet{Romero-Shaw2023MNRAS.524..245R}. Given a synthetic population containing $n$ BH-HMXB systems, the expected number of potentially observable systems is estimated as
\begin{equation}
    N_\mathrm{obs} = \sum_{i=1}^n T_{\mathrm{obs,}i} \frac{\dot M_\mathrm{SFR}}{M_\mathrm{rep}}, 
\end{equation}
where $T_{\mathrm{obs}}$ is the observable lifetime of the BH-HMXB. We define $T_\mathrm{obs}$ as the time interval during which the MS donor is between 80\% Roche-lobe filling and either fully filling its Roche lobe or evolving off the MS, whichever happens sooner. We assume a constant Milky Way star formation rate of $\dot M_\mathrm{SFR}=2 \: M_\odot \:\mathrm{yr}^{-1}$. The quantity $M_\mathrm{rep}$ denotes the total star-forming mass represented by our synthetic population. It is computed by integrating the full initial mass function and determining what fraction of that total mass is born in binaries whose primary masses fall within the range sampled in \textsc{COMPAS}, which sets the scaling between the mass of the simulated sample and that of the full stellar population it represents. In this calculation we assume a binary fraction of $f_\mathrm{bin} = 0.7$. Uncertainties on $N_\mathrm{obs}$ are estimated using bootstrap resampling of the synthetic BH-HMXB population. We generate 1000 bootstrap realizations and report the standard deviation of the resulting distribution of $N_\mathrm{obs}$. The quantity $N_\mathrm{obs}$ represents an upper bound on the number of detectable systems, as it does not account for observational selection effects such as the elevated X-ray background toward the Galactic bulge, which could prevent even a bright source like Cygnus X-1 from being confidently identified as a BH-HMXB \citep[see the discussion in][]{Romero-Shaw2023MNRAS.524..245R}.

For the BBH simulations, the metallicity of each system is drawn from a log-uniform distribution with $Z\in [0.0001,0.03]$. We use the bimodal remnant mass prescription from \citet{Maltsev2025A&A...700A..20M}, which has been previously explored with \textsc{COMPAS} in the context of BBHs \citep{Willcox2025arXiv251007573W}. We use the \texttt{BALANCED} extrapolation method for compactness-peak BHs at low metallicities as described in \citet{Willcox2025arXiv251007573W}. We vary the MS evolution prescription between \texttt{BRCEK} and \texttt{HURLEY}, and evolve $10^7$ binaries per variation.

To estimate the observable BBH merger population, we post-process the simulations using the \textsc{COMPAS} cosmic integration tool \citep{Barrett2018MNRAS.477.4685B, Neijssel2019MNRAS.490.3740N}. This framework convolves the simulated binary population with a metallicity-specific star formation history. Each simulated system is assigned a formation rate by weighting the cosmic star formation rate by the probability of forming at a given metallicity and redshift, then shifted forward in time by its binary-evolution delay time to obtain a merger rate as a function of redshift. We adopt the metallicity-dependent star formation history from \citet{vanSon2023ApJ...948..105V}. Detection rates are then obtained by applying gravitational-wave selection effects corresponding to the approximate sensitivity achieved during the LVK O3 observing run.

\begin{figure*}[t!]
    \plotone{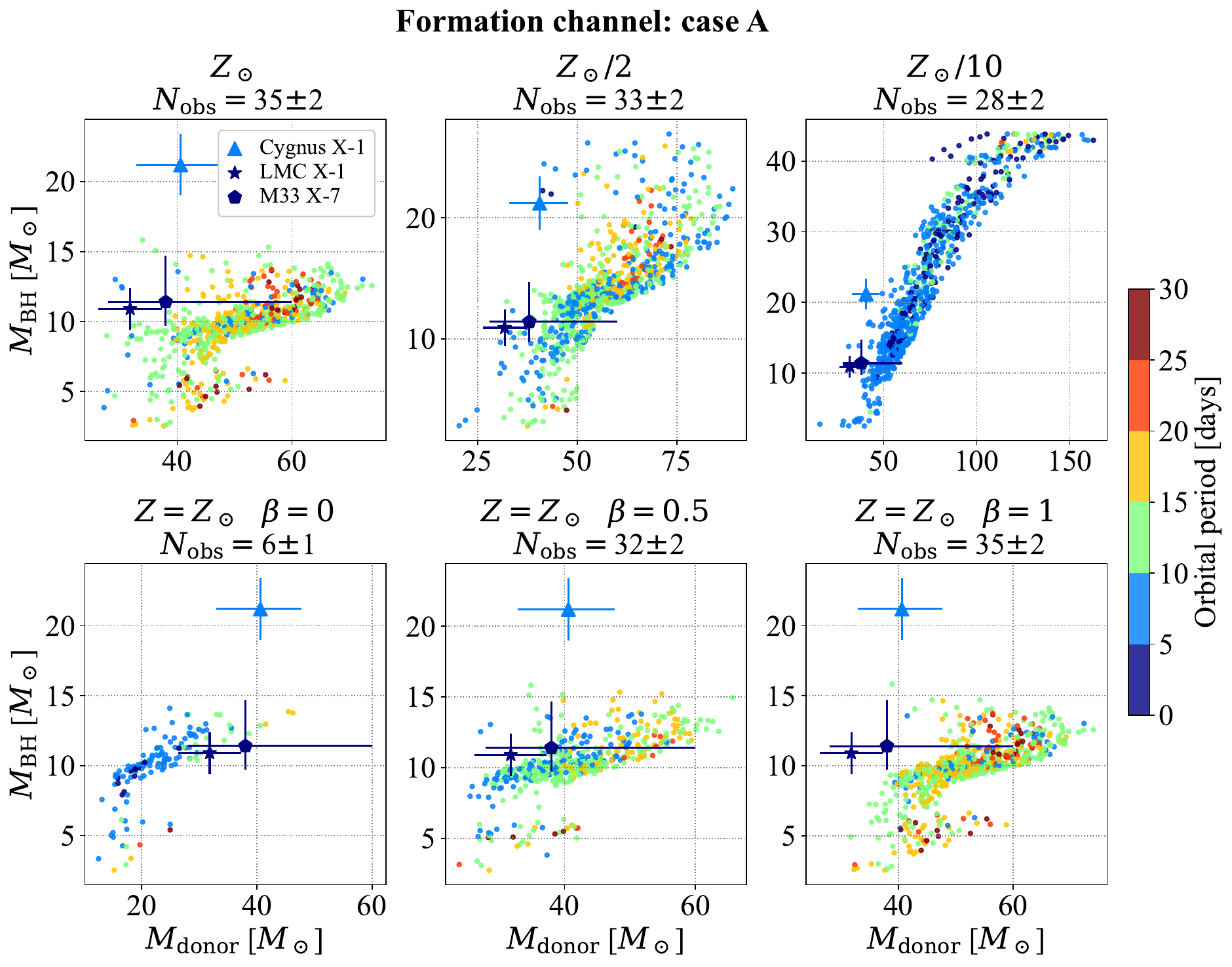}
    \caption{Mass of the BH as a function of the MS donor's mass during the HMXB phase, shown for populations at three metallicities using the default mass transfer efficiency prescription ($\beta=\beta_\mathrm{thermal}$, top panels) and using different mass accretion efficiencies at solar metallicity ($Z=Z_\odot$, bottom panels). All systems formed through case A mass transfer from the BH progenitor and are shown at the moment when the secondary fills 80\% of its Roche lobe. All points are colored based on their orbital period. Each panel also shows the predicted number $N_\mathrm{obs}$ of potentially observable BH-HMXBs, assuming a constant star formation rate of $2\:M_\odot\:\mathrm{yr}^{-1}$.   For comparison, we show the measured masses for the three confirmed BH-HMXBs: Cygnus X-1 (triangle) with the orbital period $P=5.6$ days \citep{Miller-Jones2021Sci...371.1046M}, LMC X-1 (star) with $P=3.9$ days \citep{Orosz2009ApJ...697..573O}, and M33 X-7 (pentagon) with $P=3.5$ days \citep{Ramachandran2022A&A...667A..77R}. Note that panels use different scales.}
\label{fig:caseA_channel}
\end{figure*}

\section{Results} \label{sec:results}
In this section, we present the results of our population synthesis studies.  We first examine the properties of BH-HMXBs, their dependence on metallicity and mass transfer efficiency, and the different formation channels that contribute to the population. We also explore the connection between BH-HMXBs and merging BBHs, highlighting the subset of systems in which an HMXB phase precedes BBH formation. We then turn to the population of merging double compact objects (DCOs), focusing on BBHs, where we examine their intrinsic and observable properties and the impact of the new treatment of MS evolution. Note that Sections \ref{sec:results_bh_hmxb} and \ref{sec:results_bbh} adopt different remnant mass prescriptions (see Section~\ref{sec:method} and the introduction to Section~\ref{sec:results_bbh}).

\subsection{Black hole high-mass X-ray binaries} \label{sec:results_bh_hmxb}
We begin by examining the global properties of the synthetic BH-HMXB population produced by \textsc{COMPAS}. Under the default \textsc{COMPAS} settings, simulations using the \texttt{HURLEY} MS evolution framework do not produce any BH-HMXB candidates satisfying our selection criteria (a BH with a MS companion that is at least 80\% Roche-lobe filling). This highlights the importance of accurately modeling the internal structure of massive MS stars. In the \texttt{HURLEY} formalism, the helium core mass at terminal-age main sequence (TAMS) is determined solely by the total stellar mass at TAMS, and can be significantly underestimated for stars that experience substantial MS mass loss. This suppresses the formation of sufficiently massive BHs. In contrast, the \texttt{BRCEK} MS evolution framework captures the evolution of the convective core, allowing MS stars to retain a larger helium core mass following mass loss. This enables the formation of BH-HMXB systems that are otherwise absent under the \texttt{HURLEY} framework.  We present the BH-HMXB populations separately for case A and case B/C formation channels, defined by the evolutionary phase of the BH progenitor at the onset of the first mass transfer episode.

\subsubsection{Case A mass transfer channel} \label{sec:caseA}
The case A mass transfer channel corresponds to systems in which the BH progenitor initiates mass transfer while still on the MS. This also includes systems where case A mass transfer transitions directly into case B mass transfer as the donor evolves off the MS and onto the HG (case AB mass transfer).

Figure \ref{fig:caseA_channel} shows the properties of all BH-HMXB candidates formed via stable case A mass transfer using the \texttt{BRCEK} framework. The panels show the BH mass as a function of the MS donor mass at the time when the donor becomes 80\% Roche-lobe filling, with points colored by orbital period. We use three different metallicities (top panels) and three mass accretion efficiencies at $Z_\odot$ (bottom panels). 

We compare the synthetic population against the three observed BH-HMXB systems: Cygnus X-1, LMC X-1, and M33 X-7. The metallicity of these systems is not well constrained. For Cygnus X-1, estimates range from sub-solar to super-solar (see Section~\ref{sec:discussion}), while M33 X-7 is estimated to have a metallicity of roughly half solar \citep{Magrini2007A&A...470..865M,Ramachandran2022A&A...667A..77R}, and the mean LMC metallicity is in the range of one-third to one-half solar \citep{Choudhury2016MNRAS.455.1855C}. The synthetic population spans a wide range of BH masses, with a strong dependence on metallicity.  At solar metallicity, forming systems with BH masses exceeding $\sim15\,M_\odot$ is rare, making it difficult to reproduce the configuration observed in Cygnus X-1. At lower metallicities, reduced wind mass loss leads to larger pre-collapse cores and higher BH masses, while also producing tighter systems with shorter orbital periods.

\begin{figure*}[t!]
    \plotone{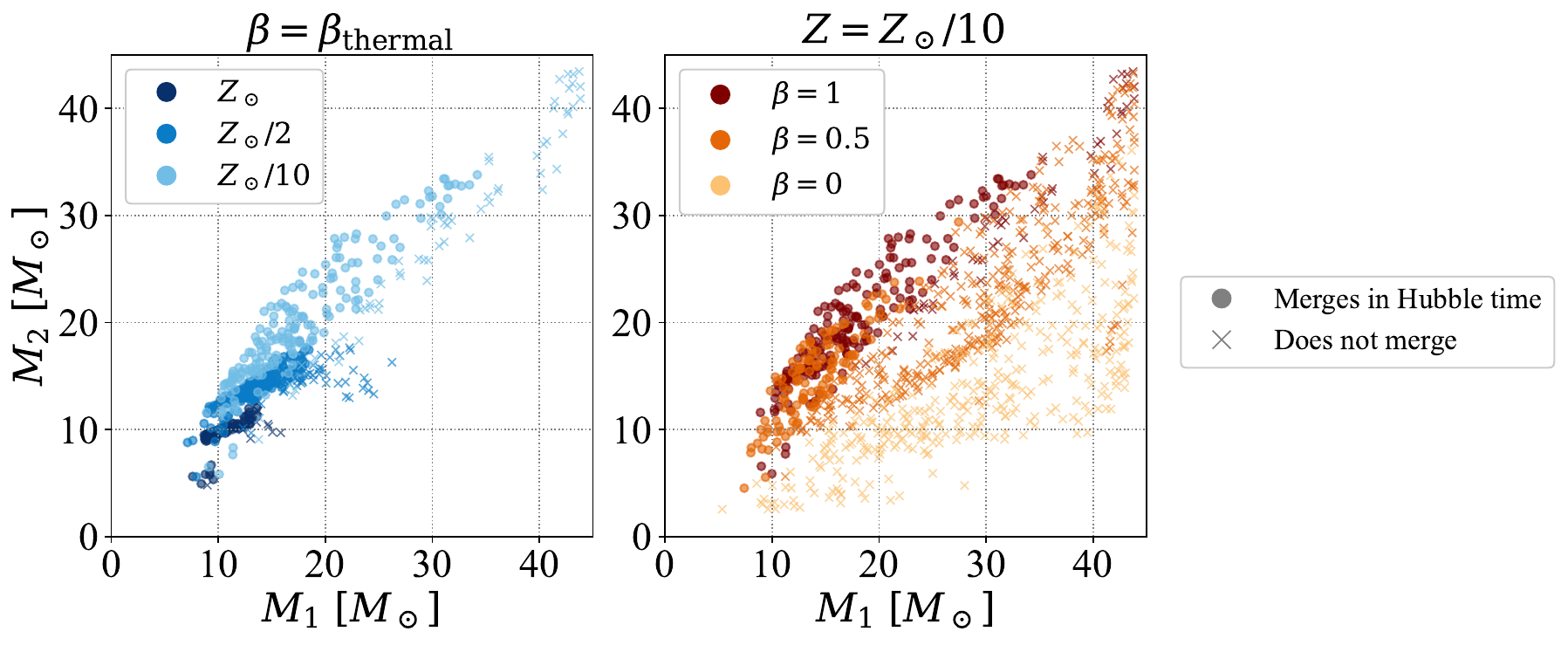}
    \caption{Primary and secondary component masses of BBHs that experienced a prior BH-HMXB phase formed through stable case A mass transfer. The left panel shows systems formed under the default mass transfer efficiency prescription based on the thermal timescale of the accretor. The right panel shows systems at fixed metallicity ($Z_\odot/10$) for varying mass transfer efficiencies $\beta$. Points represent systems that merge within a Hubble time while crosses show systems that take longer to merge.
\label{fig:BH-HMXB_merging_BBHs}}
\end{figure*}

Across all metallicities, portions of the $M_\mathrm{BH}$--$M_\mathrm{donor}$ plane remain unpopulated, corresponding to initial mass ratios that lead to unstable case A mass transfer and subsequent mergers. At solar metallicity, the distribution exhibits two distinct clusters in BH mass, one centered around $5\;M_\odot$ and another near $10\;M_\odot$. This bimodality arises from the adopted remnant mass prescription of \citet{Mandel2020MNRAS.499.3214M}, which incorporates stochasticity in the mapping between the pre-collapse carbon-oxygen core mass and final BH mass. In certain carbon-oxygen core-mass ranges, similar progenitors can produce BHs with different masses.

The bottom panels of Figure \ref{fig:caseA_channel} show the effect of varying the mass transfer efficiency $\beta$ at solar metallicity. Reducing $\beta$ lowers the donor masses of BH-HMXBs (since less material is accreted during the first mass transfer episode) and leads to lower orbital periods. More non-conservative case A mass transfer also results in a substantially smaller number of observable BH-HMXBs, consistent with the scarcity of such systems observed today, with the Milky Way expected to host only on the order of a few \citep{Romero-Shaw2023MNRAS.524..245R}. This suggests that highly conservative case A mass transfer may be disfavored for some massive binaries. In fact, the models of \citet{Lau2024ApJ...966L...7L} suggest that massive stars expand when accretion exceeds the estimate from their thermal timescale, reducing the prefactor from 10 in Eq.~(\ref{eq:thermal}) and making mass transfer less conservative.

Under the default $\beta_\mathrm{thermal}$ mass transfer efficiency prescription, which tends to be highly conservative for case A mass transfer, the number of potentially observable systems is slightly lower at lower metallicity. While the total number of BH-HMXBs formed is comparable across metallicities, the average observable lifetime is shorter at lower metallicity. This is because lower metallicity produces more massive donors, which evolve more rapidly, reducing their observable lifetime and consequently $N_\mathrm{obs}$.

The \texttt{BRCEK} framework additionally allows us to follow the surface helium and hydrogen abundances of mass-losing MS stars. The surface helium abundance becomes enhanced once the processed, helium-enriched layers are exposed due to stripping. We find that helium-enriched material (with helium mass fraction greater than 0.4) is transferred from the BH progenitor during case A mass transfer in 98\% of systems at $Z_\odot$, 93\% at $Z_\odot/2$, and 91\% at $Z_\odot / 10$. This demonstrates that case A mass transfer in massive binaries typically strips the donor down to layers within the initial mixing core boundary that contain helium-enriched material. In our models, the accreted material is assumed to have the initial hydrogen-rich composition of the donor, without accounting for the helium enrichment. However, the accretion of helium-rich material during case A mass transfer could have important consequences for the properties of the accreting star \citep{Richards2025MNRAS.544.4146R}.

To assess whether the BH-HMXB candidates would be observable as X-ray sources, we estimate the expected X-ray luminosity of the wind-accreting BHs using
\begin{equation}
    L_\mathrm{X} = \eta \, \dot{M}_\mathrm{acc} \, c^2,
    \label{eq:LX}
\end{equation}
where $\eta=0.1$ is the radiative efficiency of accretion, $c$ is the speed of light, and $\dot{M}_\mathrm{acc}$ is the Bondi-Hoyle accretion rate \citep{Bondi1944MNRAS.104..273B}, with the wind velocity assumed to be equal to the escape velocity from the donor's surface \citep{Hurley2002MNRAS.329..897H}. This estimate assumes a spherically symmetric wind from the donor. In close binaries, however, the wind acceleration can be quite anisotropic \citep{Hirai2021PASA...38...56H}, and additional effects such as X-ray feedback from the accretor onto the donor's wind are not captured. These values should therefore be taken as order-of-magnitude estimates only. For a representative system formed through the case A channel at solar metallicity,
\[
    M_\mathrm{BH}=10\:M_\odot,\:\: M_\mathrm{donor}=40\:M_\odot, \:\: P=12\:\mathrm{d},
\]
\[
     \dot{M}_\mathrm{donor}=2\times10^{-6}\:M_\odot \: \mathrm{yr}^{-1},
\]
Eq.~(\ref{eq:LX}) yields $L_\mathrm{X}\approx2\times10^{37}\:\mathrm{erg}\:\mathrm{s}^{-1}$. This is comparable to the observed X-ray luminosity of Cygnus X-1 \citep[\mbox{$\sim1$--$3\times10^{37}\:\mathrm{erg}\:\mathrm{s}^{-1}$;}][]{Vrtilek2008ApJ...678.1248V,Sugimoto2017PASJ...69...52S}.

A subset of BH-HMXBs formed in our simulations through case A mass transfer subsequently evolves into BBHs that merge within a Hubble time ($t_\mathrm{Hubble}\approx14 \:\mathrm{Gyr}$). Under the default mass transfer efficiency prescription, approximately $12\%$, 28\%, and 24\% go on to form merging BBHs at $Z_\odot$, $Z_\odot/2$, and $Z_\odot/10$, respectively, with these fractions decreasing as mass transfer becomes less conservative. Figure \ref{fig:BH-HMXB_merging_BBHs} shows the masses of BBHs that experienced an HMXB phase and formed through the case A mass transfer channel. Under the default assumptions, these BBHs preferentially have near-equal mass ratios. Adopting more non-conservative case A mass transfer broadens the mass-ratio distribution and allows the formation of systems with more uneven mass ratios. However, these unequal mass ratio systems are too wide to merge within a Hubble time. 

\subsubsection{Case B and C mass transfer channels} \label{sec:caseB_and_caseC}

\begin{figure*}[t!]
    \plotone{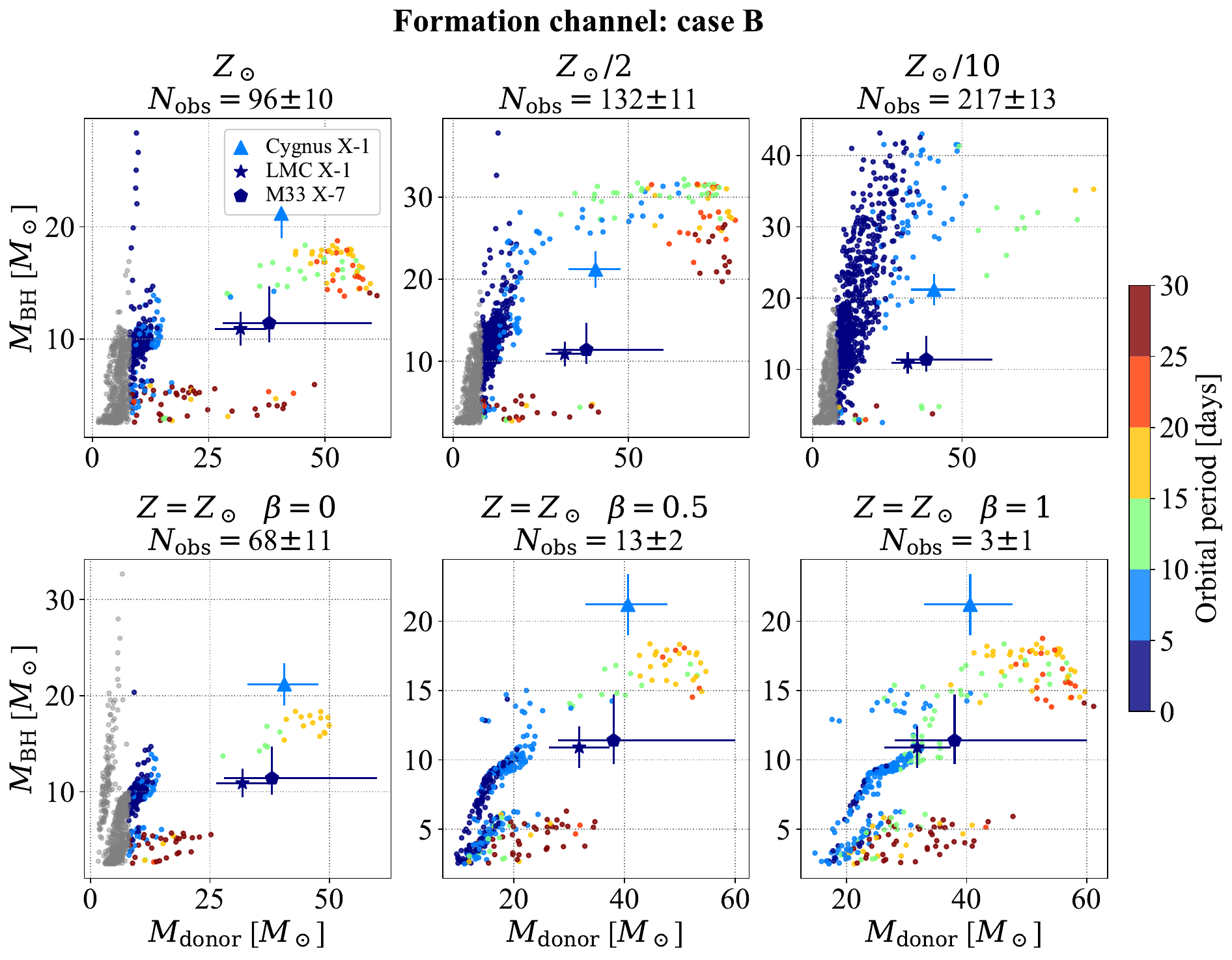}
    \caption{Same as Figure \ref{fig:caseA_channel}, but showing BH-HMXB systems formed through case B mass transfer. Systems with $M_\mathrm{donor} < 8 \: M_\odot$ are shown in gray and were excluded from the $N_\mathrm{obs}$ calculation.
\label{fig:caseB_channel}}
\end{figure*}

We next consider BH-HMXBs formed through case B mass transfer, shown in Figure \ref{fig:caseB_channel}. In this formation channel, the BH progenitor initiates its first mass transfer episode after leaving the MS. We restrict our analysis to systems undergoing stable case B mass transfer and exclude systems that enter a common-envelope phase prior to BH-HMXB formation. 

Overall, case B systems cannot reproduce the properties of observed BH-HMXBs. The predicted number of observable systems formed through case B mass transfer is highly sensitive to the assumed mass transfer efficiency, and we obtain numbers comparable to observations only in the fully conservative limit. The BHs formed through this channel are also not expected to be rapidly spinning. Since the stellar progenitors avoid interaction while on the MS, their orbits are wider and tidal locking is not established early on. By the time mass transfer does begin, the donor's core has already decoupled from its envelope, limiting the redistribution of angular momentum to the core.

In our simulations, adopting non-conservative case B mass transfer leads to a large predicted population. The majority of these systems have donor masses $M_\mathrm{donor} < 15\;M_\odot$, where the wind mass loss is weaker, and the X-ray luminosity generated by accretion is likely below the detectable threshold. Such donors also have longer MS lifetimes and spend more time near Roche-lobe filling, which substantially inflates the estimated observable rate. Systems with $M_\mathrm{donor}\lesssim8\:M_\odot$ are better classified as intermediate-mass X-ray binaries, in which stellar winds are too weak for efficient wind-fed accretion and mass transfer is instead expected to proceed via Roche-lobe overflow \citep[e.g.][]{Podsiadlowski2002ApJ...565.1107P}. Since our observability criterion from \citet{Hirai2021PASA...38...56H} is designed for wind-fed systems, we exclude donors with $M_\mathrm{donor} < 8\; M_\odot$ from the rate calculations.

We also identify a small number of BH-HMXB candidates in which the BH progenitor initiates mass transfer after core helium exhaustion (case C mass transfer).  These systems occupy a narrow region of parameter space, with $ M_\mathrm{donor}\lesssim10\:M_\odot$ and $M_\mathrm{BH}\lesssim6\: M_\odot$, and are increasingly suppressed at lower metallicity. The absence of observed counterparts may reflect inefficient wind accretion from relatively low-mass donors.

We estimate the X-ray luminosity of the case B and C systems in the same way as for case A (Eq.~\ref{eq:LX}). For a typical system formed through the case B channel at solar metallicity,
\[ 
    M_\mathrm{BH}=10\:M_\odot,\:\: M_\mathrm{donor}=14\:M_\odot, \:\:  P=5\:\mathrm{d},
\]
\[
     \dot{M}_\mathrm{donor}=5\times10^{-8}\:M_\odot \: \mathrm{yr}^{-1},
\]
which yields $L_\mathrm{X}\approx2\times10^{36}\:\mathrm{erg}\:\mathrm{s}^{-1}$. This is an order of magnitude fainter than the estimated X-ray luminosity of Cygnus X-1 \citep{Vrtilek2008ApJ...678.1248V,Sugimoto2017PASJ...69...52S}. A typical case C system with 
\[ 
    M_\mathrm{BH}\approx3\:M_\odot,\:\: M_\mathrm{donor}\approx6\:M_\odot, \:\: P\approx2\:\mathrm{d}, 
\]
\[
     \dot{M}_\mathrm{donor}=2\times10^{-10}\:M_\odot \: \mathrm{yr}^{-1},
\]
yields $L_\mathrm{X}\approx3\times10^{33}\:\mathrm{erg}\:\mathrm{s}^{-1}$, an order of magnitude below the faint end of the observed persistent HMXB luminosity function in the Milky Way \citep[$\sim 10^{34}$ erg s$^{-1}$;][]{Lutovinov2013MNRAS.431..327L}. While case C systems may thus be genuinely unobservable, the overabundance of case B systems produced in the models points to model limitations, likely in the treatment of mass transfer (see Section \ref{sec:discussion}).

\subsection{Binary black holes} \label{sec:results_bbh}

We next explore the impact of the new MS evolution framework on the population of double compact objects (DCOs), which represent the end products of massive binary evolution. Note that for these simulations, we adopt the remnant mass prescription from \citet{Maltsev2025A&A...700A..20M}, which provides updated explodability criteria for the neutrino-driven supernova mechanism. Its impact on the BBH population was explored by \citet{Willcox2025arXiv251007573W}, who find that it produces a chirp-mass distribution in better agreement with gravitational-wave observations than the default prescription. We therefore consider it more appropriate for the BBH population analysis, while retaining the \citet{Mandel2020MNRAS.499.3214M} prescription for the BH-HMXB simulations for consistency with previous studies \citep[e.g.][]{Romero-Shaw2023MNRAS.524..245R}.  A subset of BBHs in our models form via chemically homogeneous evolution, in which rotationally induced mixing maintains a near-uniform composition throughout the star \citep{Maeder1987}. We exclude BBHs formed through this channel from our analysis, since it remains subject to various uncertainties and contributes primarily to the higher-mass end of the BBH population \citep{Marchant:2016,MandelDeMink:2016,Riley2021}.

\subsubsection{Intrinsic properties}

\begin{figure}[t!]
    \plotone{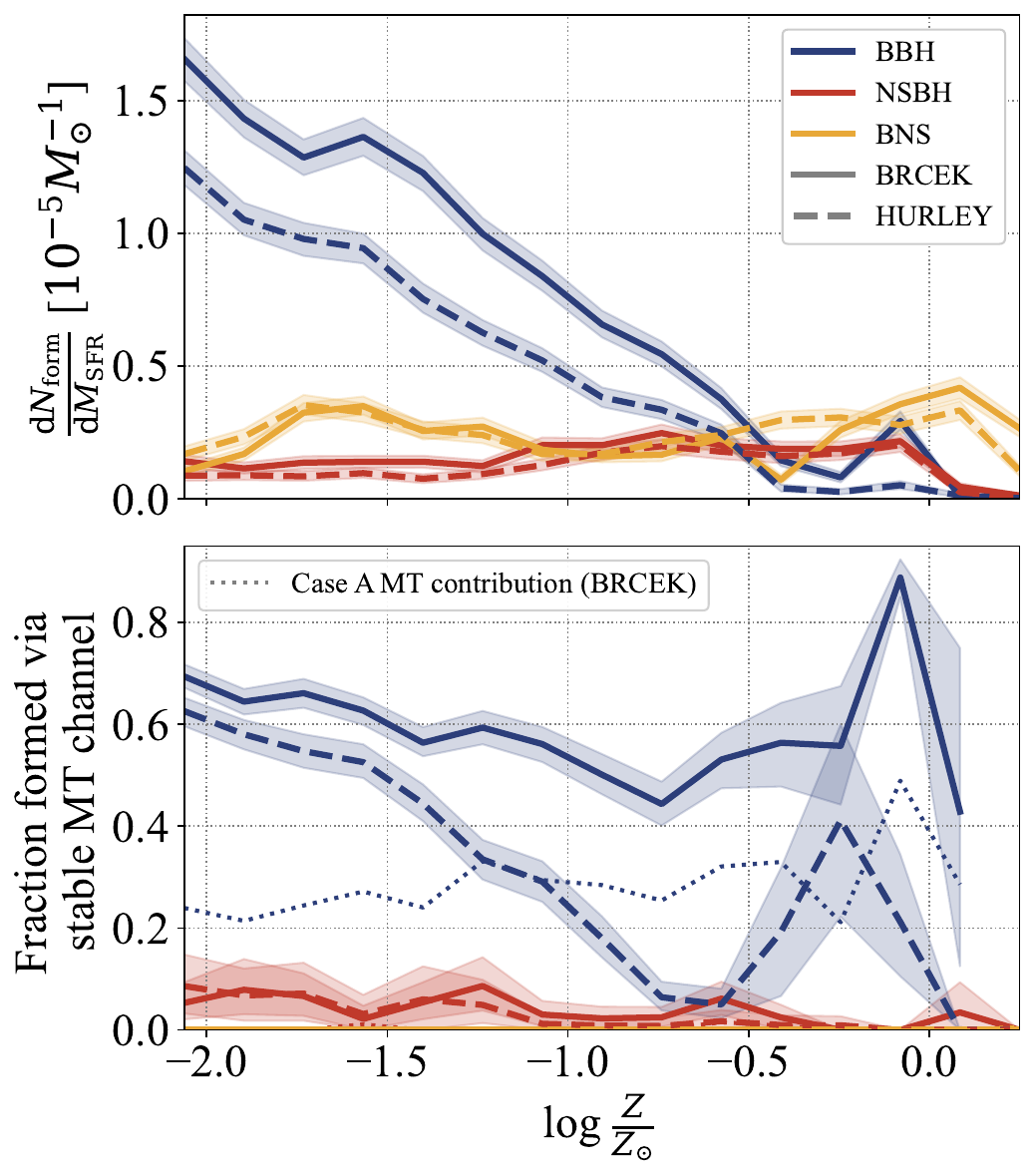}
    \caption{Intrinsic yields of merging double compact objects per unit star-forming mass as a function of metallicity (top panel), and the relative contribution from the SMT channel (bottom panel). BBH systems are shown in blue, NSBH in red, and BNS in yellow. Solid and dashed lines show results using the \texttt{BRCEK} and \texttt{HURLEY} MS evolution frameworks, respectively. The dotted line in the bottom panel shows the contribution of case A mass transfer to the SMT channel under the \texttt{BRCEK} framework. Shaded areas represent 90\% confidence regions, obtained using non-parametric bootstrapping.
\label{fig:intrinsic_yields}}
\end{figure}

Figure~\ref{fig:intrinsic_yields} shows the intrinsic yields of merging BBHs, neutron star–black hole (NSBH) binaries, and binary neutron stars (BNS) per unit star-forming mass as a function of metallicity (top panel). Figure~\ref{fig:intrinsic_yields} also shows the fraction of DCOs formed through the SMT channel (bottom panel). We define the SMT channel as formation of a DCO without experiencing a CE episode and excluding the chemically homogeneous evolution channel. The CE channel then means that the system experienced at least one CE event prior to DCO formation. We only consider binaries that merge within a Hubble time. 

For systems containing neutron stars (BNS and NSBH), we find negligible differences between the \texttt{BRCEK} and \texttt{HURLEY} frameworks, as expected. These systems form primarily through the CE channel and do not experience mass transfer from MS donors. Consequently, modifications to the MS evolution have limited impact on the formation efficiency of BNS and NSBH systems. NS progenitors also originate primarily from stars with initial masses $\lesssim20\,M_\odot$, where wind-driven mass loss during the MS evolution  plays a smaller role in determining the final carbon-oxygen core mass. 

In contrast, the BBH yield is enhanced across the full metallicity range under the \texttt{BRCEK} prescription. This reflects a systematic underestimation of the helium core mass at TAMS in the \texttt{HURLEY} framework for stars that have undergone substantial mass loss.  This enhancement is not uniform, and it features a localized peak in the BBH yield at $Z\approx0.7\:Z_\odot$ that is much weaker under \texttt{HURLEY}. This peak arises from the remnant mass prescription of \citet{Maltsev2025A&A...700A..20M}, in which there exists a narrow range of carbon-oxygen core masses at around $7\: M_\odot$ for which the star undergoes complete fallback and the BH forms without a natal kick. At $Z\approx0.7\:Z_\odot$, Wolf-Rayet wind mass loss limits the final carbon-oxygen core mass, so a wide range of progenitors end up with carbon-oxygen core masses in this narrow range. As a result, fewer binaries are disrupted by the supernova kick, and more of them survive to contribute to the merging BBH yield.

The bottom panel of Figure~\ref{fig:intrinsic_yields} shows the fraction of BBHs formed through the SMT channel. Under \texttt{BRCEK}, this fraction remains fairly constant at around 60\% across most of the metallicity range, rising to $\sim80$\% at sub-solar metallicities. The dotted line shows that case A mass transfer accounts for around half of all systems formed through the SMT channel. Note that we only show the case A contribution for \texttt{BRCEK}, as the \texttt{HURLEY} framework produces only a small number of merging BBHs that underwent case A mass transfer ($<0.3\%$ of the merging BBH population). Under \texttt{HURLEY}, case A mass transfer tends to produce BBHs that are too wide to merge within a Hubble time because stellar radii of partially stripped stars are overestimated, leading to excessive mass transfer and binary widening.

\begin{figure}[t!]
    \plotone{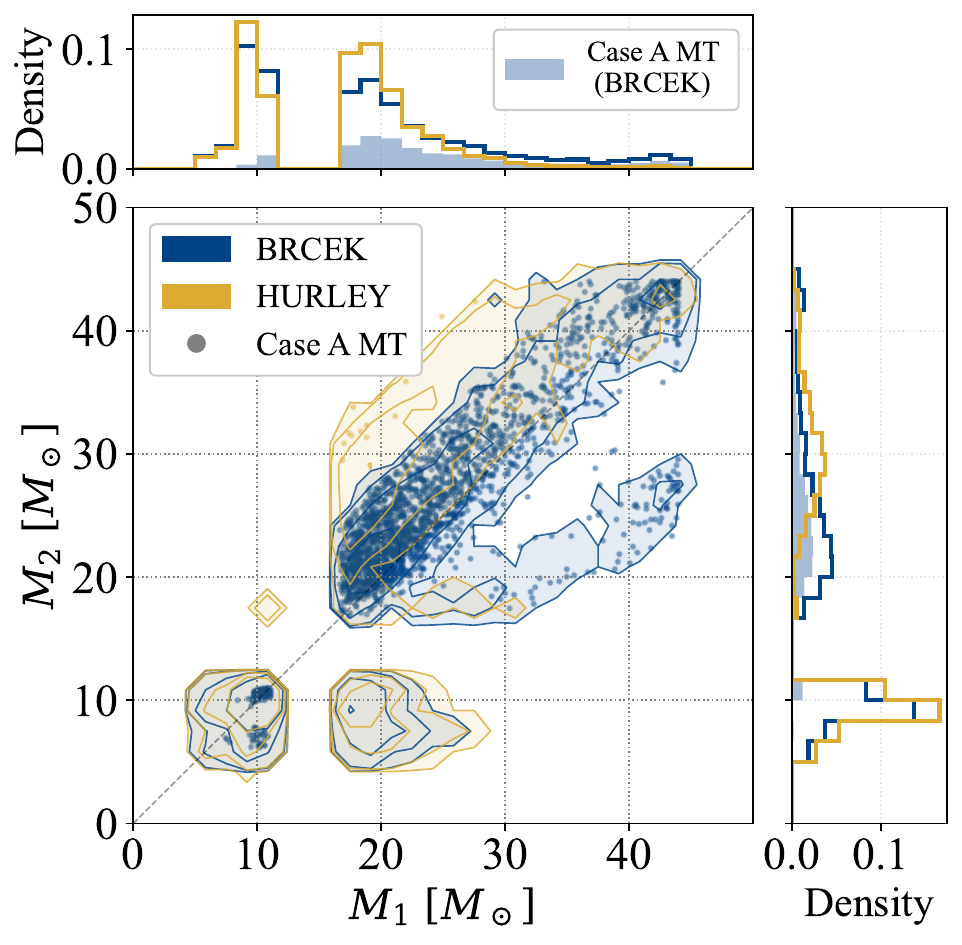}
    \caption{Mass distribution of merging BBHs for the \texttt{BRCEK} (blue contours) and \texttt{HURLEY} (yellow contours) MS evolution frameworks. The top and right panels show the normalized $M_1$ and $M_2$ distributions, respectively. Blue and yellow points in the main panel and blue shaded histograms in the marginal panels show the subset of systems that formed through the case A mass transfer channel. Very few such systems ($<0.3\%$ of the full merging BBH population) form under \texttt{HURLEY}, so we show the case A contribution in the histograms only for the \texttt{BRCEK} framework.
\label{fig:intrinsic_mass_distributions}}
\end{figure}

We then explore the masses of the merging BBH population, shown in Figure \ref{fig:intrinsic_mass_distributions}. The remnant mass prescription from \citet{Maltsev2025A&A...700A..20M} does not produce BHs with masses $\sim15 \:M_\odot$, which appear as a gap in the mass distributions. Systems formed through different channels occupy distinct regions of parameter space. The overdensity at $M_1\approx20 \:M_\odot$ and $M_2\approx10 \:M_\odot$ arises purely from CE evolution. Systems with $M_1\gtrsim 20 \:M_\odot$ and $M_2\gtrsim 20 \: M_\odot$ form almost exclusively through SMT, and the overdensity at $M_1,M_2\approx10\: M_\odot$ results from a combination of both channels. The majority of systems (55\% under \texttt{BRCEK} and 57\% under \texttt{HURLEY}) lie above the diagonal $M_1=M_2$ line, indicating that mass-ratio reversal, where the initially less massive star ends up forming the more massive BH, plays a significant role in shaping the BBH population. This is mainly a consequence of stable mass transfer, in which the primary transfers a substantial fraction of its mass to the secondary before collapsing, leaving the secondary as the more massive star.

An interesting feature emerging under the \texttt{BRCEK} framework is a branch of massive systems that extends below the diagonal line. This arises from the new treatment of rejuvenation in MS accretors \citep{Brcek2026ApJ..1002...78B}. These systems begin as massive binaries with close to equal mass ratios. The primary fills its Roche lobe as a MS or HG star, while the secondary is close to the end of its own MS lifetime. Under \texttt{BRCEK}, rejuvenation is much less significant for accretors close to core hydrogen exhaustion. This then allows the secondary to complete its MS evolution and subsequently transfer mass back onto the primary while the latter is still a naked helium star. Under default \textsc{COMPAS} assumptions, this second mass transfer episode is highly conservative, resulting in a significant mass growth of the primary. In reality, this episode may be less conservative or may lead to the re-expansion of the primary and further mass loss, and much of the accreted mass may not ultimately contribute to the BH mass of the primary. We therefore interpret these systems with some caution. 

Two further differences between the \texttt{BRCEK} and \texttt{HURLEY} frameworks are visible in Figure \ref{fig:intrinsic_mass_distributions}. First, massive SMT systems formed under \texttt{BRCEK} tend to have mass ratios closer to unity, with the exception of the uncertain branch below the diagonal line. Second, the high-mass peak in the $M_2$ histogram is shifted toward lower masses. Both effects stem from how case A and B mass transfer are treated. Under \texttt{BRCEK}, the mass transfer rate slows once processed, helium-enriched layers become exposed at the donor's surface. The rate is unaffected under \texttt{HURLEY}, producing more extreme mass ratios at the end of case A mass transfer. Additionally, the post-MS envelope mass can remain a substantial fraction of the total stellar mass under \texttt{HURLEY}, while under \texttt{BRCEK} it is typically limited to 10\% of the total mass. This restricts how much mass can be transferred onto the secondary during case B mass transfer, shifting the BBH $M_2$ peak to lower masses.

We estimate the intrinsic local merger rate at redshift $z=0$ by convolving the simulated BBH population with the cosmic star formation rate and a redshift-dependent metallicity distribution as explained in Section \ref{sec:method}. We obtain the following local BBH merger rates
\[
    \mathcal{R}_\mathrm{BRCEK} = (33 \pm 2) \: \mathrm{Gpc}^{-3}\:\mathrm{yr}^{-1},
\]
\[
    \mathcal{R}_\mathrm{HURLEY} = (10 \pm 1) \: \mathrm{Gpc}^{-3}\:\mathrm{yr}^{-1},
\]
where the uncertainties represent the 90\% confidence interval derived by bootstrapping over the simulated population, capturing only the sampling noise from the finite number of simulated systems.

\subsubsection{Observable properties}

Because BBHs dominate the population of DCOs observed through gravitational waves, we further investigate whether these structural differences leave an observable imprint on the mass-ratio and chirp-mass distributions, following the cosmic-integration procedure described in Section~\ref{sec:method}. When constructing the cumulative distribution functions (CDFs), each simulated merging BBH is weighted by its detection rate, so the resulting distribution reflects the observable population.

\begin{figure*}[t!]
    \plotone{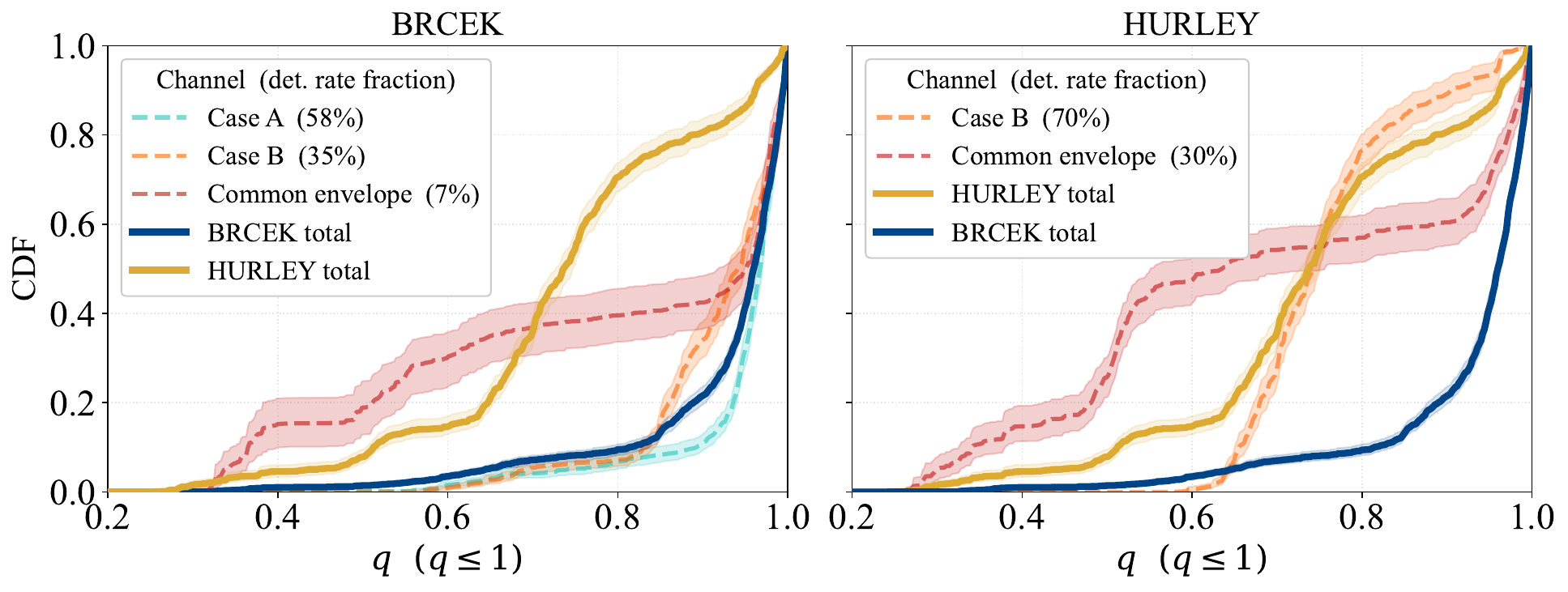}
    \caption{Cumulative distribution of the mass ratio $q$ for detectable BBH mergers under the \texttt{BRCEK} (left) and \texttt{HURLEY} (right) MS evolution frameworks. Colored dashed lines show the contribution from case A (cyan), case B (orange), and CE (red) formation channels with the legend indicating each channel's fraction of the total detection rate. Thick solid lines show the total detectable population for the  \texttt{BRCEK} (blue) and \texttt{HURLEY} (yellow) frameworks, repeated in both panels for comparison. Shaded regions show 90\% confidence intervals from non-parametric bootstrapping. 
\label{fig:mass_ratio_CDF}}
\end{figure*}

The mass-ratio CDF is shown in Figure \ref{fig:mass_ratio_CDF}, together with the contribution from each formation channel. The SMT channel is split into case A and case B, based on the first mass transfer episode. Under \texttt{HURLEY}, there is a sharp rise in the CDF around $q\approx0.7$, driven by the case B channel. Under \texttt{BRCEK}, the case A channel instead makes a significant contribution to the detectable population, and the overall distribution shows a strong preference for near-equal mass ratios. In both frameworks, the majority of the detection-weighted population consists of mass-ratio reversal systems (61\% of the detectable BBH population under \texttt{BRCEK} and 74\% under \texttt{HURLEY}).
This preference arises because \texttt{BRCEK} provides a mechanism to slow radial expansion once the helium-enriched layers become exposed at the donor's surface, reducing the amount of mass transferred during case A mass transfer. Under \texttt{HURLEY}, the same systems instead experience substantial mass transfer after mass-ratio reversal, driving their mass ratios toward lower values.

\begin{figure}[b!]
    \plotone{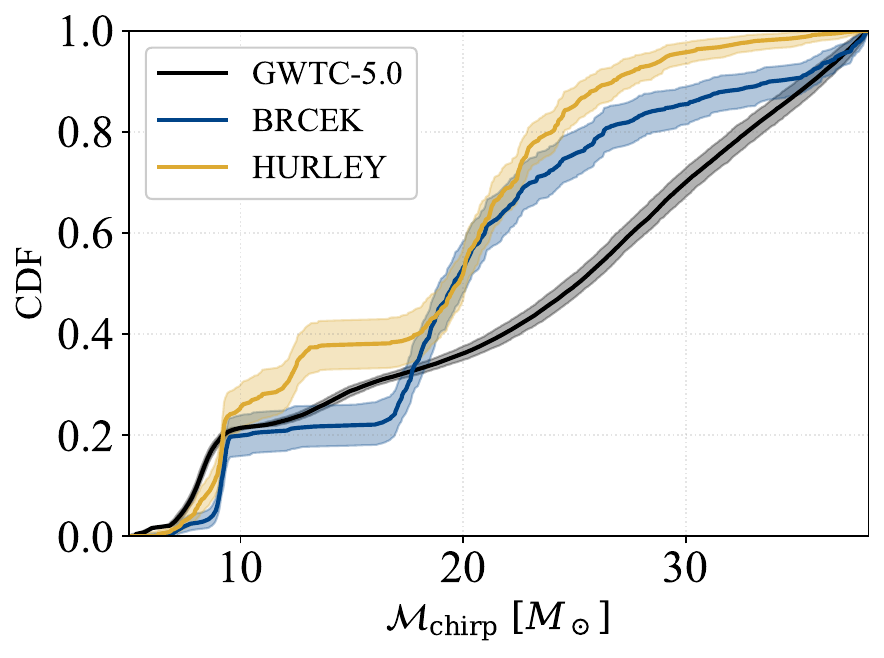}
    \caption{Cumulative chirp-mass distribution of detectable BBH mergers predicted by the \texttt{BRCEK} (blue) and \texttt{HURLEY} (yellow) MS evolution frameworks, compared to detected BBH candidates from GWTC-5.0 (black). Shaded regions show 90\% confidence intervals.
\label{fig:chirp_mass}}
\end{figure}

The chirp mass is a particularly well-measured parameter in gravitational-wave observations, defined as 
\begin{equation}
\mathcal{M}_\mathrm{chirp} = \frac{(M_1 M_2)^{3/5}}{(M_1+M_2)^{1/5}}, 
\end{equation}
where $M_1$ and $M_2$ are the BBH component masses. The chirp-mass CDF is shown in Figure~\ref{fig:chirp_mass}, comparing the \texttt{BRCEK} and \texttt{HURLEY} frameworks with GWTC-5.0 \citep{LIGO2026arXiv260527223T}. We select observed BBHs as systems with $M_1,M_2 > 3\:M_\odot$, false alarm rate $< 1$ yr$^{-1}$, and parameter $p_\mathrm{astro}>0.5$, yielding a sample of 249 BBH merger events. The uncertainty band around the observed distribution is constructed by drawing one chirp-mass posterior sample per event, creating a CDF with the drawn values, and repeating this process to build a 90\% confidence interval. For the simulated populations, we draw 249 systems from the models with a probability proportional to their detection rate and repeat the process to obtain the corresponding 90\% confidence interval.

The bimodal remnant mass prescription gives rise to three corresponding features in the chirp-mass distribution: a low-mass peak around $8\:M_\odot$, a second peak around $12\:M_\odot$, and a broader feature extending from $\sim17\:M_\odot$ to $\sim 30 \:M_\odot$. The first feature consists of a combination of systems formed through SMT and CE, while the second and third arise predominantly from the CE and SMT channels, respectively. 

There are distinct differences between the two frameworks. The second feature, which arises from a combination of a higher-mass and a lower-mass BH, is suppressed under \texttt{BRCEK} relative to \texttt{HURLEY}. This follows from the new framework's preference for near-equal mass ratios (Figure \ref{fig:mass_ratio_CDF}) and the reduced contribution from the CE channel. Although the overall incidence of CE events is comparable between the two frameworks, their outcomes differ considerably for systems that interact through case AB mass transfer. Because the reduced mass transfer responsible for near-equal mass ratios also leaves these systems on tighter post-mass-transfer orbits, they are less likely to survive the subsequent CE episode without merging. Combined with the larger number of BBHs formed via the SMT channel under \texttt{BRCEK}, this suppresses the fractional contribution of the CE channel to the detectable population. We find that the CE formation channel only contributes to chirp masses up to $\sim14\:M_\odot$, and all BBH systems with larger chirp masses come from the SMT channel.

At $\mathcal{M}_\mathrm{chirp}\gtrsim18\:M_\odot$, the \texttt{BRCEK} framework produces more systems than \texttt{HURLEY}, including some with $\mathcal{M}_\mathrm{chirp}>35\:M_\odot$ that are absent under \texttt{HURLEY}. The maximum BH mass is $\sim 45 \:M_\odot$ in our models and is set by the pulsational pair-instability prescription \citep{Marchant2019ApJ...882...36M,Stevenson2019ApJ...882..121S}. BBHs exceeding this limit do exist in our simulations on the upper side of the pair-instability mass gap but form exclusively at very low metallicities through the chemically homogeneous channel, which we have excluded from this analysis.

\section{Discussion} \label{sec:discussion}
\subsection{Black hole high-mass X-ray binaries}
The inability to reproduce the mass of Cygnus X-1 at solar metallicity can have several possible explanations. First, its metallicity could be lower than solar. As an interacting system, the original metallicity of Cygnus X-1 is difficult to constrain observationally, with some studies suggesting a sub-solar value based on another system in the same association \citep{Daflon2001ApJ...552..309D} and others finding it to be super-solar \citep{Shimanskii2012ARep...56..741S}. Second, the BH mass itself could be lower than reported by \citet{Miller-Jones2021Sci...371.1046M}. \citet{Ramachandran2025A&A...698A..37R} estimate a BH mass of $12.7$--$17.8\,M_\odot$, which is reproducible by our models even at solar metallicity.

A third scenario is that Cygnus X-1 is currently observed after an earlier episode of fast case A mass transfer, rather than just before Roche-lobe filling \citep{Podsiadlowski2003MNRAS.341..385P}. If the donor remains close to filling its Roche lobe after the fast mass transfer phase, the system would spend considerably longer in this configuration than in the brief window prior to Roche-lobe overflow. We explore this possibility in Appendix~\ref{sec:Appendix} using detailed stellar evolution models. Following the fast mass transfer phase, the stripped donor in these models remains close to Roche-lobe filling, with masses, orbital period, donor's radius, and surface helium abundance consistent with those observed in Cygnus X-1. However, the donor in our models continues to overflow its Roche lobe with a mass transfer rate between $10^{-6}$ and  $10^{-5}\:M_\odot\:\mathrm{yr}^{-1}$, which would power accretion onto the BH near the Eddington limit, in tension with the observed X-ray luminosity \citep{Vrtilek2008ApJ...678.1248V}. For this scenario to be consistent with observations, the donor would instead need to be slightly detached from its Roche lobe, with the BH fed by the stellar wind, similar to the post-interaction picture of \citet{Xing2025A&A...693A..27X}.  
Irradiation-driven stellar expansion during the phase of rapid mass transfer could lead to subsequent donor contraction and the cessation of Roche-lobe overflow.  Alternatively, if Cygnus X-1 is really in Roche-lobe overflow today, the accretion rate onto the BH must be significantly lower than the mass transfer rate from the donor, indicating that a future outburst is imminent.  Moreover, we find only a small number of such post-interaction systems in our population synthesis models. This is because mass transfer from a more massive MS donor onto a less massive BH is unstable at $q\lesssim0.5$ under the default mass transfer stability prescription in \textsc{COMPAS}, preventing most systems from reaching this phase. Further improvements to the estimates of the radius of partially stripped stars could help assess how common such systems are.

If BH-HMXBs are intrinsically more likely to be observed after the mass transfer episode from the MS donor onto the BH accretor, rather than just before it, this has direct implications for the implied mass transfer efficiency. A donor observed post-mass-transfer must have been more massive prior to being stripped than its currently observed mass suggests. This implies that the secondary accreted a non-negligible amount of mass during the initial mass transfer episode, in tension with the strongly non-conservative case A mass transfer favored in our models. Disentangling the stability and efficiency of case A mass transfer from the resulting observability of the BH-HMXB system would be a valuable direction for future work.

Several physical mechanisms could nonetheless drive case A mass transfer to be more non-conservative than commonly assumed. Non-conservativeness may originate from spin-up of the accretor during the initial mass transfer phase. Once the accretor is spun up beyond critical rotation, further accretion may be suppressed. However, rotationally limited accretion might be disfavored for some observed binary systems \citep[e.g.][]{Vinciguerra2020MNRAS.498.4705V,Lechien2025ApJ...990L..51L}. Additionally, we simplify the treatment of stellar winds during Roche-lobe overflow, and wind mass loss is not considered when the donor transfers mass on a nuclear timescale. This effect may be further enhanced once the donor is stripped beyond its initial convective core boundary, exposing helium-rich material at the surface and boosting wind mass loss. Observational evidence supports the possibility of highly non-conservative case A mass transfer, including studies of binaries hosting Wolf-Rayet stars \citep{Petrovic2005,Schurmann2025A&A...704A.219S}, double-lined eclipsing binaries \citep{deMink2007}, and Algols \citep{Sen2026ApJ..1000....2S}. Observations also suggest that HMXBs are more numerous in low-metallicity environments \citep{Douna2015A&A...579A..44D}, a trend our models reproduce only under non-conservative case A mass transfer, providing more evidence that mass transfer in massive binaries may be more non-conservative than commonly assumed.

Our models also predict a large number of BH-HMXBs for which there is no observational counterpart, particularly those formed through the case B channel. As shown in Section \ref{sec:caseB_and_caseC}, many of the case B and C systems would produce relatively low X-ray luminosities, making them difficult to detect. The predicted number of systems through this channel also depends sensitively on the assumed mass transfer physics. In our simulations, we assume that the mass lost during non-conservative mass transfer carries away the specific angular momentum of the accretor. Under this assumption, adopting non-conservative case B mass transfer leads to overestimated populations. If the mass was instead lost through the $L_2$ Lagrange point, it would carry away substantially more specific angular momentum than under our isotropic re-emission assumption. This additional angular momentum loss could drive more systems into merger, naturally suppressing the number of BH-HMXBs formed through this channel.

The fraction of BH-HMXBs formed through case A mass transfer that go on to form merging BBHs in our models is consistent with some previous studies \citep{Fishbach2022ApJ...929L..26F,Gallegos-Garcia2022ApJ...938L..19G,Romero-Shaw2023MNRAS.524..245R}, though the fraction could be even lower \citep[e.g.][]{Liotine2023ApJ...946....4L}. We additionally explored the impact of the remnant mass prescription by applying the \citet{Maltsev2025A&A...700A..20M} prescription to the BH-HMXB population, finding that the fraction of systems contributing to the merging BBH population drops to $\sim1\%$ at $Z_\odot$, compared to $\sim12\%$ under the \citet{Mandel2020MNRAS.499.3214M} prescription. This highlights that uncertainties in the remnant mass prescription can be as important as mass transfer assumptions when studying the connection between BH-HMXBs and gravitational-wave sources.

\subsection{Binary black holes}

Many previous studies of the SMT channel for BBH mergers report a peak in the mass-ratio distribution at $q\approx0.7$ \citep{vanSon2022ApJ...931...17V,Broekgaarden2022ApJ...938...45B,Dorozsmai2024MNRAS.530.3706D}. With the new treatment of MS evolution, we instead find a strong preference for $q\approx1$, consistent with some recent studies \citep{Klencki2026A&A...706A.296K,Briel2026arXiv260203629B}. This result also broadly agrees with the recent GWTC-5.0 release, where the data shows reduced evidence for the previously reported peak at $q\approx0.7$ \citep{LIGO2026arXiv260527226T}. Furthermore, the mass distribution shown in Figure \ref{fig:intrinsic_mass_distributions} reveals distinct subpopulations that are broadly consistent with those identified in a recent population-level analysis of GWTC-5.0 \citep{Guttman2026arXiv260722011G}. We attribute the difference in the mass-ratio peak to the treatment of case A and B mass transfer. Under \texttt{HURLEY}, the amount of mass transfer is overestimated, leading to more extreme mass ratios following the interaction. This effect is further amplified once the donor leaves the MS, since \texttt{HURLEY} continues to attribute a large fraction of the total stellar mass to the envelope while underestimating the helium core mass.

Case A mass transfer as a channel for producing rapidly spinning BHs is interesting to consider. Individual gravitational-wave events with high primary spin and mass ratio not close to unity, such as GW241011 and GW241110 \citep{Abac2025ApJ...993L..21A}, would be challenging to produce with our models. Effectively all merging BBHs that experienced case A mass transfer have $q>0.5$. Our models also produce a population of systems in which case A or B mass transfer from the primary onto the secondary is later followed by mass transfer from the secondary back onto the now-stripped, naked helium star primary. Even a modest amount of accreted mass could be sufficient to spin it up to near-critical rotation, potentially providing an additional channel for producing rapidly spinning BHs at more extreme mass ratios. However, it is not clear what the outcome of such mass transfer episodes should be. Future studies could help explain whether such interactions provide a viable channel for spinning BBH formation.
 
We find that the predicted local merger rates of BBHs are also affected by the choice of MS evolution framework. \citet{LIGO2026arXiv260527226T} report the BBH merger rate at $z\approx0.2$ in the range of 27.5--49.4 Gpc$^{-3}$ yr$^{-1}$, consistent with the rate obtained using the \texttt{BRCEK} framework. It is also important to emphasize that the predicted rates depend on many other assumptions \citep[e.g.][]{Willcox2025arXiv251007573W}, and we only model the isolated binary evolution channel of BBH formation (for reviews, see \citealt{Mapelli:2021, MandelFarmer:2018, Mandel2022LRR....25....1M}).

We find that the SMT channel dominates the detectable BBH population.  However, the relative contribution of the CE channel depends strongly on additional assumptions \citep[e.g.,][]{Broekgaarden2026arXiv260605322B}, including the prescription for angular momentum loss during non-conservative mass transfer, the adopted critical mass ratios for stability, and the choice of CE parameters. Notably, the suppressed CE contribution under \texttt{BRCEK} largely removes the intermediate chirp-mass mode identified by \citet{Willcox2025arXiv251007573W} in their trimodal chirp-mass distribution, which arises from pairing a low-mass and a high-mass BH. In our models, this mode is populated by the CE-formed $M_1\approx20\:M_\odot,\:M_2\approx10\:M_\odot$ overdensity (Figure~\ref{fig:intrinsic_mass_distributions}), which is strongly suppressed under \texttt{BRCEK}. Future work systematically exploring these uncertainties under the new MS evolution framework in the context of BH formation would help clarify their relative importance.

\section{Conclusion}            \label{sec:conclusion}
In this work, we use the rapid binary population synthesis code \textsc{COMPAS}, incorporating a new treatment of MS evolution that tracks the convective core under mass loss or accretion. Massive binaries commonly interact while their components are still on the MS, and an accurate treatment of this phase is essential for correctly modeling the masses and orbital properties of the BH populations they may form. Our results demonstrate that this is indeed the case across both BH populations we examined.

For BH-HMXBs, we find that the more detailed treatment of MS evolution in the \texttt{BRCEK} framework is necessary to reproduce the high BH masses of observed systems. Matching the orbital periods, donor masses, and observed rates favors case A mass transfer that is more non-conservative than assumed by the default prescription. We also find that the transfer of helium-enriched material during case A mass transfer is common among massive binaries, which could have implications for the structure and subsequent evolution of the accreting star.

For merging BBHs, the new MS evolution framework alters the predicted rates as well as the chirp-mass and mass-ratio distributions of the detectable merger population. We find that case A mass transfer makes a significant contribution to the SMT channel for BBH formation, further highlighting the importance of detailed MS evolution models in binary population synthesis. Systems formed through the SMT channel show a strong preference for near-equal mass ratios. This shifts the predicted population away from the previously reported mass-ratio peak at $q\approx0.7$ and toward $q \approx1$, in better agreement with the most recent gravitational-wave population results from GWTC-5.0. The suppression of the $q \approx 0.7$ peak, together with the enhanced contribution of the SMT channel and enhanced local merger rates, illustrates how changes to the treatment of MS evolution can propagate into observable differences in the gravitational-wave population.

Together, these results highlight that an accurate treatment of MS evolution in binary population synthesis can alter our predictions for both electromagnetic and gravitational-wave observables. Future work should explore how additional assumptions further shape these populations. This includes the stability criteria for case A and case B mass transfer, the treatment of angular momentum loss during non-conservative mass transfer, and the structural consequences of helium-enriched accretion. As gravitational-wave catalogs continue to grow and additional BH-HMXBs are discovered, these observations will provide increasingly powerful tests of the MS evolution physics underlying both populations. More direct tests may come from spectroscopic and photometric observations of massive binaries currently undergoing, or having recently completed, case A mass transfer. In particular, measurements of surface abundances, mass-radius relations, and orbital period distributions could probe the predictions of the \texttt{BRCEK} framework directly.

\section*{Data availability}
\textsc{COMPAS} simulations used in this work are publicly available at GW Landscape (\url{https://gwlandscape.org.au/publications/}). \textsc{MESA} inlists and history files used in Appendix~\ref{sec:Appendix} are available on Zenodo \citep{brcek_2026_22005805}.

\begin{acknowledgments}
We acknowledge support from the Australian Research Council (ARC) Centre of Excellence for Gravitational Wave Discovery (OzGrav), through project number CE230100016. A.B. acknowledges support from the Commonwealth through an Australian Government Research Training Program Scholarship [DOI: \url{https://doi.org/10.82133/C42F-K220}]. R.H. acknowledges support from the RIKEN Special Postdoctoral Researcher Program for junior scientists. J.Y.C. acknowledges support from Monash University through the Summer and Winter Vacation Research Scholarship Program.
\end{acknowledgments}

\appendix
\section{Cygnus X-1 as a post-interaction system} \label{sec:Appendix}
The donor in Cygnus X-1 is observed close to filling its Roche lobe, a configuration that may be explained by assuming a coincidental observation just prior to the onset of mass transfer. Here we explore the alternative that Cygnus X-1 is instead being observed after an initial episode of fast case A mass transfer from the massive donor onto the BH. We investigate this possibility using the stellar evolution code \textsc{MESA} version r25.12.1 \citep{Paxton2011,Paxton2013,Paxton2015,Paxton2018,Paxton2019,Jermyn2023ApJS..265...15J}, simulating only the mass transfer from a MS star onto the BH in order to test whether this scenario can reproduce the observed properties of Cygnus X-1.

\begin{figure*}[t!]
    \plotone{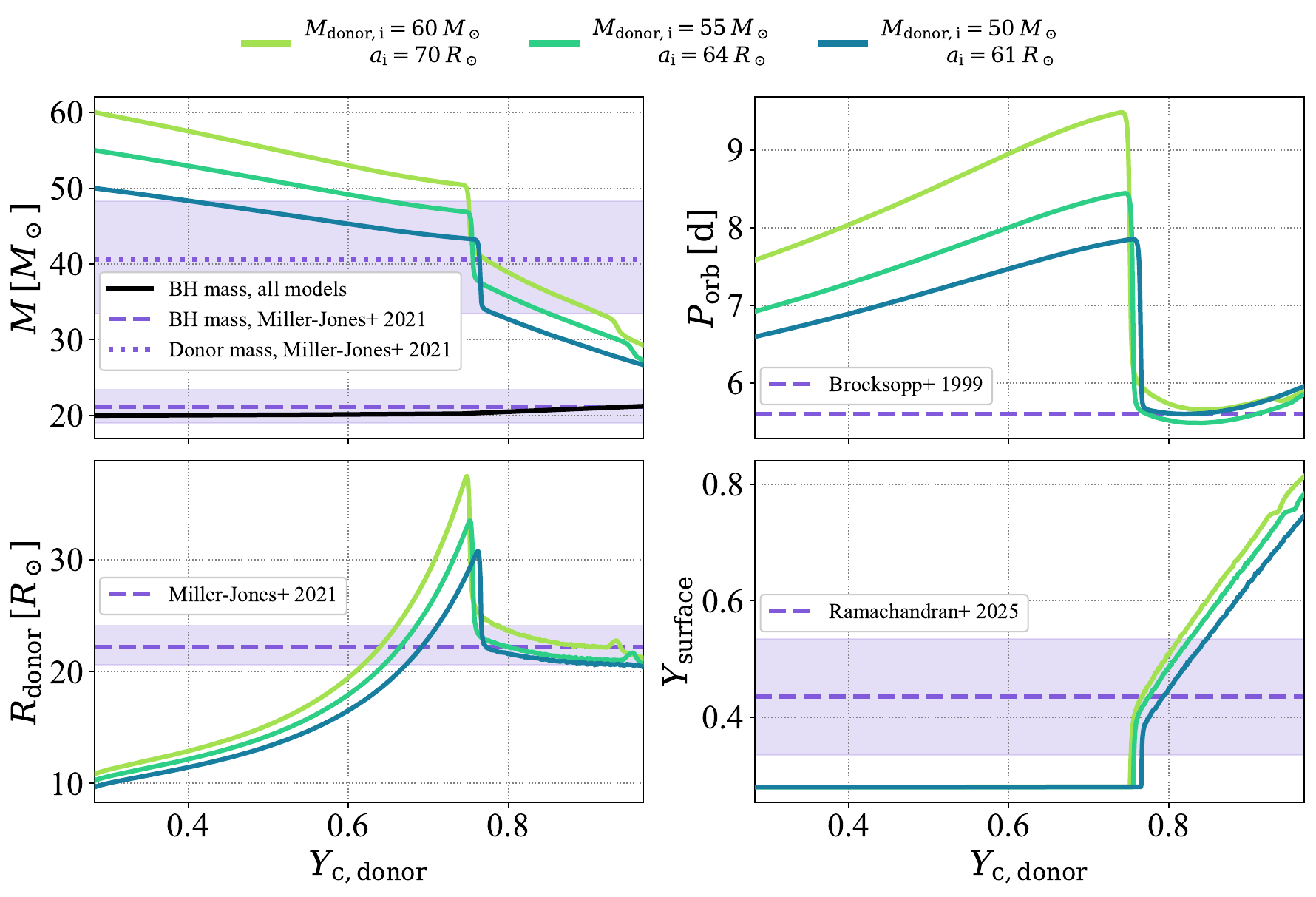}
    \caption{Evolution of the donor mass and BH mass (top left), orbital period (top right), donor radius (bottom left), and surface helium abundance (bottom right) as a function of the donor's central helium fraction $Y_\mathrm{c}$, for three \textsc{MESA} models with different initial donor masses $M_\mathrm{donor,i}$ and separations $a_i$ (see top legend). Purple horizontal lines and shaded regions show the observed values and their uncertainties for Cygnus X-1 from \citet{Miller-Jones2021Sci...371.1046M} (donor and BH mass, donor radius), \citet{Brocksopp1999A&A...343..861B} (orbital period), and \citet{Ramachandran2025A&A...698A..37R} (surface helium abundance). The sharp features in the donor mass, radius, and orbital period correspond to the onset of mass transfer and the transition from fast to slow case A mass transfer.
\label{fig:CygX1_mesa_models}}
\end{figure*}

We consider three initial configurations with varying MS donor masses and orbital separations, summarized in Figure~\ref{fig:CygX1_mesa_models}. In all cases, we adopt solar metallicity and an initial BH mass of $20 \: M_\odot$. We initialize the evolution with a MS donor and a BH, bypassing the uncertainties associated with the prior binary evolution. The pre-interaction donor mass of $\sim45\:M_\odot$ and orbital period of $\sim9\:\mathrm{d}$ in our models are broadly consistent with systems formed through case A mass transfer in \textsc{COMPAS} (see Figure~\ref{fig:caseA_channel}). We assume Eddington-limited accretion onto the BH, and treat angular momentum loss during non-conservative mass transfer using isotropic re-emission.

Figure~\ref{fig:CygX1_mesa_models} shows that the considered models are able to reproduce the observed masses of the donor and the BH, the orbital period, and the donor's radius, as well as its elevated surface helium abundance. In our models, the transition from fast to slow mass transfer occurs once the donor has been stripped down to its helium-rich layers. Such stars have qualitatively different chemical structures compared to non-interacting stars, leading to mass-radius relations that differ from those in our \textsc{COMPAS} models. During this slow phase, the donor transfers mass at a rate of $\sim10^{-6}$--$10^{-5}\:M_\odot\:\mathrm{yr}^{-1}$. This is up to an order of magnitude above the Eddington accretion rate of the BH.  If the BH accreted at the Eddington limit, the X-ray luminosity would be well above that observed for Cygnus X-1 \citep{Vrtilek2008ApJ...678.1248V}.

One caveat is that our \textsc{MESA} models initialize the donor as a MS star with primordial composition, whereas our population synthesis results suggest that the accretor likely accreted helium-enriched material during the initial case A mass transfer from the BH progenitor (Section~\ref{sec:caseA}). Such enrichment could alter the internal structure, mixing properties, and radius evolution of the donor, potentially affecting the subsequent evolution and the parameter space of matching models. 

Overall, these models illustrate that a post-interaction configuration can reproduce several key properties of Cygnus X-1. However, the mass transfer rate during the slow phase remains in tension with the observed X-ray luminosity, raising an important open question.


\bibliography{references}{}
\bibliographystyle{aasjournalv7}



\end{document}